\documentclass[structabstract]{aa}  
\usepackage{graphicx}
\usepackage{txfonts}
\usepackage{natbib}
  \def\simge{\mathrel{\raise1.16pt\hbox{$>$}\kern-7.0pt
    \lower3.06pt\hbox{{$\scriptstyle \sim$}}}}           
  \def\simle{\mathrel{\raise1.16pt\hbox{$<$}\kern-7.0pt
    \lower3.06pt\hbox{{$\scriptstyle \sim$}}}}           
\def\fm{\hbox{$.\!\!^{\rm m}$}}

\newcommand{\pv}{\ensuremath{P_V}}
\newcommand{\nv}{\ensuremath{N_V}}

\newcommand{\bz}{\ensuremath{\langle B_z\rangle}}

\newcommand{\nz}{\ensuremath{\langle N_z\rangle}}

\begin{document}
\title{The magnetic fields of hot subdwarf stars\thanks{Based on
    observations collected at the European Organisation for
    Astronomical Research in the Southern Hemisphere, Chile under
    observing programmes 072.D-0290 and 075.D-0352, or obtained from
    the ESO/ST-ECF Science Archive Facility.}}


   \author{J. D. Landstreet
          \inst{1,2}
          \and
          S. Bagnulo
          \inst{1}
          \and
          L. Fossati
          \inst{3}
          \and
          S. Jordan
          \inst{4}
          \and
          S. J. O'Toole
          \inst{5}
          }

   \institute{Armagh Observatory, College Hill, Armagh, BT61 9DG,
              Northern Ireland, United Kingdom.
              \email{jls@arm.ac.uk; sba@arm.ac.uk}
              \and
              Department of Physics \& Astronomy, University of Western 
              Ontario, London, ON N6A 3K7, Canada.
              \email{jlandstr@uwo.ca}
              \and
              The Open University, Walton Hall, Milton Keynes, MK7 6AA,
              United Kingdom.
              \email{l.fossati@open.ac.uk}
              \and
              Astronomisches Rechen-Institut, Zentrum f\"ur Astronomie der Universit\"at Heidelberg, 
              M\"onchhofstr. 12-14, D-69120 Heidelberg, Germany.
              \email{jordan@ari.uni-heidelberg.de}
              \and
              Australian Astronomical Observatory, PO Box 296, Epping NSW 1710, Australia.
              \email{otoole@aao.gov.au}
             }

   \date{Received 23 February 2012, accepted 29 March 2012}

 
  \abstract
  {Detection of magnetic fields has been reported in several sdO and
    sdB stars. Recent literature has cast doubts on the reliability of
    most of these detections. The situation concerning the occurrence
    and frequency of magnetic fields in hot subdwarfs is at best
    confused.}
  {We revisit data previously published in the literature, and we
    present new observations to clarify the question of how common
    magnetic fields are in subdwarf stars.}
  {We consider a sample of about 40 hot subdwarf stars. About 30 of
    them have been observed with the FORS1 and FORS2 instruments of
    the ESO VLT.  Results have been published for only about half of
    the hot subdwarfs observed with FORS.  Here we present new
    FORS1 field measurements for 17 stars, 14 of which have never been
    observed for magnetic fields before.  We also critically review
    the measurements already published in the literature, and in
    particular we try to explain why previous papers based on the same
    FORS1 data have reported contradictory results.}
  {All new and re-reduced measurements obtained with FORS1 are shown
    to be consistent with non-detection of magnetic fields. We explain
    previous spurious field detections from data obtained with FORS1
    as due to a non-optimal method of wavelength calibration.  Field
    detections in other surveys are found to be uncertain or doubtful,
    and certainly in need of confirmation.}
   { There is presently no strong evidence for the occurrence of a
     magnetic field in any sdB or sdO star, with typical
     longitudinal field uncertainties of the order of 2--400~G. It
     appears that globally simple fields of more than about 1 or 2~kG in
     strength occur in at most a few percent of hot subdwarfs, and may
     be completely absent at this strength. Further high-precision
     surveys, both with high-resolution spectropolarimeters and with
     instruments similar to FORS1 on large telescopes, would be very
     valuable.}


   \maketitle
%

\section{Introduction}
 
Hot subdwarf (sdB and sdO) stars are subluminous relative to
main sequence stars of the same effective temperature. They are the
most common type of faint blue star in the galaxy.

The sdB stars have effective temperature $T_{\rm eff} \la 40\,000$\,K,
and generally have H-rich atmospheres. They are mostly deficient in
atmospheric He, and often in several other light elements such as C,
O, Mg and Al. A few of them show greatly enriched iron peak element
abundances. The sdO stars have a large range of H/He abundance ratios,
and show a variety of compositions. Their $T_{\rm eff}$ values
mostly lie between 40\,000 and 90\,000~K.  

Both sdB and sdO stars have lost most or all of their H-rich
outer envelopes, due to mass loss and/or binary mass transfer, and are
now burning He in their cores.  In turn, a small fraction of white
dwarfs will descend from them. However, the details of how stars
evolve to become hot subdwarfs, and particularly how such a wide variety
of atmospheric chemistries is produced, are very poorly understood.

Because many hot subdwarfs show strong atmospheric composition anomalies,
which on the main sequence are frequently found in magnetic Ap-Bp
stars, it is natural to wonder if some or all hot subdwarfs also possess
detectable magnetic fields. The presence or absence of a field could
be an underlying parameter influencing the observed surface
composition, as in magnetic upper main sequence stars.
If fields can be detected in hot subdwarfs, these stars
could complement the main sequence magnetic stars as laboratories in
which to study the operation of such processes as atomic diffusion,
surface convection, internal mixing, and mass loss in the presence of
global magnetic fields \citep{land05}.

More generally, strong global fields are known to occur in some hot
stars (i.e. stars without deep outer convection zones and active
current magnetic dynamos) on the main sequence and in the white dwarf
state.  Hot subdwarfs represent a possible intermediate stage
between the main sequence and the collapsed white dwarf state, in
which the interior of the former main sequence star is exposed to
observation, and again lacks deep outer convection.  Detection of a
magnetic field in any hot subdwarf could be very helpful to
understanding the evolution of a global internal magnetic field as the
host star changes in structure due to stellar evolution. Even clear
evidence that fields are {\it not} present in hot subdwarfs down to
some small upper limit provides a useful constraint on theory.

For these reasons, several surveys have been conducted to search for
magnetic fields in sdB and sdO stars. At first, it seemed that these
surveys were very successful in detecting kG-strength fields in a 
number of hot subdwarfs \citep{elkin96,otooleetal05}, but recent
work by \citet{petitetal11} and \citet{bagnuloetal12}, has raised
doubts about the reliability of most of the detections. A recent
survey of cool  sdB stars by \citet{mathysetal12}, based on FORS2 data,
reported a suspected field in only one out of ten  sdB stars.
Currently the situation is quite confused; it is not clear if magnetic
fields are common among hot subdwarfs, or rare, or even if any have
been detected at all. If any hot subdwarf fields have in fact been
detected, almost nothing is known about the characteristics of the
fields.

A first step is to clearly establish what is known from previous work
in this field. The goal of this paper is to clarify the present
situation, to establish which field detections (if any) are robust,
and which are doubtful. To do this, we review the previous surveys,
expanding on work already presented by \citet{bagnuloetal12}.
Furthermore, we publish for the first time further field measurements
of  17 hot subdwarfs carried out with FORS1.  Based on a sample
of 41 stars, we finally draw conclusions on the incidence of magnetic
fields in hot subdwarfs.

\section{Previous magnetic measurements}\label{Sect_Previous}

A number of searches for magnetic fields in hot subdwarf stars have been
carried out, for a total (to the best of our knowledge) of 25 objects
surveyed.

\citet{borraetal83} used a Cassegrain filter polarimeter equipped with
narrow band interference filter to measure the circular polarisation
in the wings of H$\beta$. They were able to obtain one field
measurement each of \object{Feige~86} = BD$+$30~2431 and
\object{HD\,149382} = BD$-03$~3967, but because of the faintness of
the stars ($V \sim 9 - 10$) and the use of only a single spectral
line, the standard errors of measurement reported were about 2800~G.
No significant field was detected to this precision in either star.

Spectropolarimetric observations of the sdO stars \object{BD$+$75~325}
and \object{BD$+$25~2534} by \citet{elkin96} were obtained using a
classical fixed polarisation analyser on the Russian 6-m telescope to
measure Zeeman  polarisation in the He~{\sc i} D3 line $\lambda$~5876. Two
hot subdwarfs, BD$+$75~325 and Feige~66 = BD$+$25~2534, were observed,
and field detections at the kG level were reported for both stars. It
is difficult to know how secure these detections are. Although
\citet{elkin96} standardised his measurements each night with
observations of both a null and a polarised standard star, repeated
field measurements of one of the two hot subdwarfs on a single night could
vary by as much as 1\,kG, which is of the same order as the reported
fields.  The reported standard errors of 250~G or more for the
polarisation standard (53\,Cam), with about a dozen sharper spectral
lines in the 120\,\AA--wide window used, suggests that the uncertainty
of field measurement using only one broader line could be of the order
of 1\,kG. Our view is that these reported detections certainly would
need to be confirmed by further observations with a more sensitive 
method before they could be considered secure.

\citet{otooleetal05} reported field measurements of six different 
sdB and sdO stars ($T_{\rm eff}$ between 25\,000 and 70\,000 K), all
obtained during a single night. Each star was observed once using
FORS1 in spectropolarimetric mode, and in each of the six stars a
field of about $-1$\,kG was detected, with reported uncertainties of
the order of 100--230\,G. 

\citet{petitetal11} recently published new ESPaDOnS observations of
two hot subdwarfs, Feige 66, and HD 76431, in which field detections had
been previously reported by \citet{elkin96}, and by
\citet{otooleetal05} respectively. No fields were detected with the new
observations, with \bz\ uncertainties of 200\,G and 55\,G,
respectively. Furthermore, \citet{petitetal11} re-analysed the old
FORS1 data of HD 76431 used by \citet{otooleetal05}, and the
revised field measurement was found consistent with zero. In
conclusion, the reported field detections for Feige 66 and HD76431 are
both quite uncertain.

All FORS1 measurements by \citet{otooleetal05} have been re-analysed
by \citet{bagnuloetal12}.  For all six stars, the new reductions
failed to confirm the reported fields. It appears to us that the
problem with the original reductions by \citet{otooleetal05} is that
separate wavelength calibrations were used for the two waveplate
settings of the FORS1 polarimeter, thus effectively sabotaging the
possibility of using measurements from the two settings to cancel out
first-order errors in the relative wavelength calibration of the two
analysed beams. As discussed by \citet{bagnuloetal09}, tiny numerical
differences in the wavelength calibration of the frames obtained at
different position angles of the retarder waveplate may lead to
noticeable spurious polarisation signals. In the present case, the
residual uncertainty in the different calibrations, of order
0.03\,\AA, is quite large enough to introduce field measurement errors
of the order of 1\,kG, and this might be the reason all six stars
observed appear to have very similar longitudinal field strengths.
Thus  \citet{bagnuloetal12} concluded that the field detections
reported by \citet{otooleetal05} were spurious. A similar problem was
discovered for FORS1 observations of central stars of planetary
nebulae, leading to revision of previously reported field detections
(Jordan et al., submitted to A\&A).

Three field measurements of the sdO star \object{WD~1036+433} =
Feige~34 were reported by \citet{valyavinetal06}, using a
low-resolution spectropolarimeter at the prime focus of the Russian
6-m telescope.  This instrument is conceptually rather similar to
FORS1, with a resolving power of $\sim 2000$, and a rotating
quarter-wave plate in the polarisation analyser, thus making it
possible to cancel out errors in the wavelength calibration of the two
beams to first order.  One of the three observations showed a magnetic
field of $\langle B_z \rangle = 9.6 \pm 2.6$~kG, significant at the
$3.7\,\sigma$ level The other two measurements were both consistent
with zero field, although one was significant at the $2\,\sigma$
level. We consider that this field detection may be correct, but
because  only one measurement is significant, and only at a
little more than the $3\,\sigma$ level, it certainly requires further
confirming observations.

Recently, \cite{savanovetal11} analysed the spectropolarimetric
observations of the bright ($B=11\fm 8$) non-radially oscillating sdB
star \object{Balloon 90100001} = \object{TYC 2248-1751-1} taken with
the main stellar spectrograph of the 6\,m Special Astrophysical
Observatory. They  did not detect any magnetic field, and their
  measured field value was $34\,\pm 63$\,G, considerably lower than
the field strengths reported for other hot subdwarfs by
\cite{otooleetal05} and \citet{valyavinetal06}.

During the course of a survey of about 60 DA white dwarfs for weak
fields, made with the Steward Observatory CCD spectropolarimeter,
field measurements were obtained for four sdB stars by
\citet{kawkaetal07}. This survey is remarkable for the number of
non-detections reported, which strongly confirm that the instrument is
not prone to false positive detections. No fields were found in the
observations (one or two per star) of the four hot subdwarfs, with
reported uncertainties in the range of 4--12~kG.

Finally, a new survey of 10 cool sdB stars using FORS2 (which now has the
polarimetric optics formerly installed on FORS1) has just been
reported by \citet{mathysetal12}. This group standardised their
observations with measurements of the magnetic Ap star HD~142070. Each
star was observed between two and four times. One observation (of
three) of the star SB~290 is significant at about the $4\sigma$ level,
and a second is significant at about the $2.5\sigma$ level. It is
quite possible that a field has been detected in this star, but, as
the authors point out,  their provisional detection certainly
needs to be confirmed by further observations.

What is remarkable about the observations discussed above is the fact
that, although apparently significant fields have been reported in a
total of 10 hot subdwarfs, new measurements of only two of these stars
have been reported \citep[by][]{petitetal11} which  were able to
confirm (or disprove, in this specific case) the reality of the
reported field detections.

\section{New FORS1 magnetic field measurements}

\begin{table*}
\caption{\label{Tab_Observations}
Fundamental stellar parameters and FORS1 magnetic field measurements of hot subdwarf stars.
}
\begin{center}
\begin{tabular}{lllrrrrrr@{$\,\pm\,$}lc}
\hline
\hline
\multicolumn{2}{l}{STAR} &     
Spect.      &              
$T_{\rm eff}$ &              
$\log g$    &              
Exp. time   &              
Peak SNR    &              
\multicolumn{1}{c}{MJD} &  
\multicolumn{2}{c}{\bz} &     
                     \\    
\multicolumn{2}{l}{} &       
type        &                    
(K)         &                    
            &                    
(s)         &                    
(\AA$^{-1}$) &                    
            &                    
\multicolumn{2}{c}{(G)} &   
                             \\  
\hline
\object{CD-38 222}   & SB\,290       & SD:B  & 28200 & 5.5 & 3024 & 1865 & 53574.364 &$   32 $& 150  &\\ 
                     &               &       &       &     & 1512 & 1430 & 53624.097 &$ -172 $& 202  &\\
\object{HD 4539}     &PG\,0044$+$097 & SD:B  & 27000 & 5.5 & 1260 & 1340 & 53058.218 &$  554 $& 181  &\\
\object{PHL 932}     & PG\,0057$+$155& SD:B  & 33650 & 5.7 & 2760 &  960 & 53593.256 &$   -3 $& 307  &\\
\object{PG 0133$+$144}                                    
                     &Baloon 92627001& SD:B  &       &     & 1824 &  775 & 53638.250 &$-1074 $& 388  &\\
\object{CD-24 731}   & SB\,707       & SD:B  & 37000 & 6.0 & 2700 & 1205 & 53629.135 &$  363 $& 411  &\\
\object{PG\,0342$+$026}                                    
                     &WD\,0342$+$026 &SD:B   & 26200 & 5.7 & 2634 & 1570 & 53593.377 &$ -113 $& 172  &\\
\object{HD 127493}   & BD$-$22 3804  & SD:O  & 41000 & 5.1 &  984 & 1165 & 53571.047 &$  232 $& 178  &\\
\object{ALS\,9313}   &LS\,IV\,$-12$\,1& SD:O & 60000 & 4.5 & 2760 & 1000 & 53566.068 &$  207 $& 239  &\\
\object{HD 149382}   & BD$-$03 3967  & SD:OB & 34200 & 5.9 &  696 &  490 & 53458.390 &$ -268 $& 731  &\\
\object{HD 171858}   &CD$-$23 14565  & SD:B  &       &     &  876 & 1945 & 53512.357 &$  -32 $& 136  &\\
\object{CD-51 11879} & LSE\,263      & SD:O  & 70000 & 4.9 & 2766 & 1270 & 53512.395 &$  392 $& 262  &\\
\object{HD 188112}   &CD$-$28 16258  & SD:B+? & 21500 & 5.7 & 1170 &  470 & 53565.291 &$ -228 $& 690  &\\
\object{CD-23 15853} & LSE\,21       & SD:O  & 100000 &     & 2580 & 1020 & 53533.347 &$ -332 $& 561  &\\
\object{HD 205805}   &CD$-$46 14026  & SD:B  & 25000 & 5.0 & 1152 & 1810 & 53533.384 &$ -139 $& 120  &\\
\object{JL\,87}      &EC 21435-7634  & SD:B  & 28000 & 5.2 & 2760 &  950 & 53597.196 &$  -64 $& 153  &\\
\object{WD 2148$+$286}                                     
                     & BD$+$28 4211  & SD:O  &       &     & 1572 & 1240 & 53533.414 &$ -121 $& 415  &\\
\object{CD-35 15910} & SB\,814       & SD:B  & 28800 & 5.4 & 4800 & 1385 & 53598.378 &$  246 $& 232  &\\[2mm]
\object{TD1 32702}   &[CW83] 0512-08 & SD:B  & 38000 & 5.6 & 2640 & 1600 & 53058.025 &$  209 $& 160  &*\\
\object{CPD-64 481}  & PPM 354969    & SD:B  & 27500 & 5.0 & 2640 & 1350 & 53058.069 &$  100 $& 230  &*\\
\object{CD-31 4800}  & ALS\,591      & SD:O  & 44000 & 5.4 & 1440 & 1565 & 53058.215 &$   90 $& 140  &*\\
\object{HD 76431}    &BD$+$02\,2100  & SD:B  & 31000 & 4.5 & 1800 & 2675 & 53058.255 &$   38 $&  76  &*\\
\object{PG\,0909$+$275}                                    
                     &PG\,0909$+$276 & SD:B  & 35400 & 6.0 & 7200 & 1345 & 53058.139 &$  -23 $& 194  &*\\
\object{CD-22 9142}  & EC 11481-2303 & SD:O  & 42000 & 5.8 & 1510 & 1236 & 53134.112 &$  636 $& 416  & **\\ 
                     &               &       &       &     & 1510 & 848  & 53144.110 &$  348 $& 708  & **\\ 
\object{CD-46 8926}  & LSE\,153      & SD:O  & 70000 & 4.8 & 9900 & 2275 & 53058.347 &$  -68 $& 158  &*\\[2mm]
\hline
\end{tabular}
\end{center}
\noindent
In the last column, one asterisk (*) means that the observations were previously published by \citet{otooleetal05}; two asterisks (**), that the data were published by \citet{jordanetal07}.

\end{table*}
\begin{figure}
\rotatebox{270}{\scalebox{0.34}{
\includegraphics*[0.0cm,0.8cm][21cm,28cm]{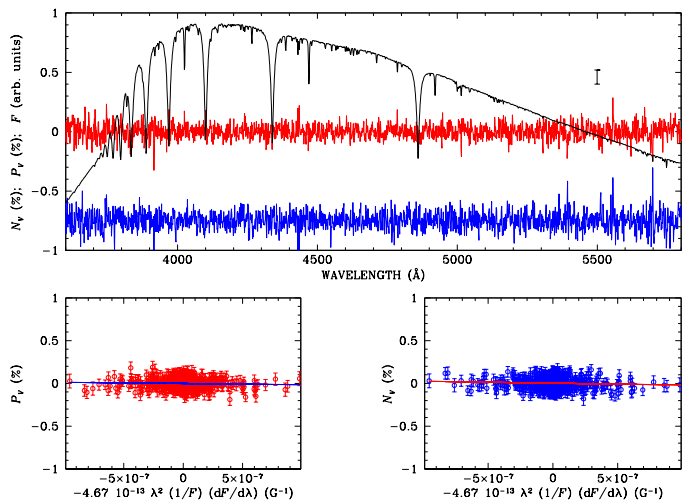}}}\\
\caption{\label{Fig_HD205805} The observations of HD\,205805 obtained
  with FORS1 on 2005-06-12.  The top panel shows the observed flux $F$
  (black solid line, in arbitrary units, and not corrected for the
  instrument response), the $\pv = V/I$ profile (red solid line
  centred about 0), and the null profile \nv\ (blue solid line, offset
  by $-0.75$\,\% for display purpose). The null profile is expected to
  be centred about zero and scattered according to a gaussian with
  $\sigma$ given by the \pv\ error bars.  A typical \pv\ error bar is
  shown in the upper right of the upper panel.  The slope of the
  interpolating lines in the bottom panels give the mean longitudinal
  field from \pv\ (left bottom panel) and from the null profile (right
  bottom panel), both calculated using the H Balmer and metal lines.
  The corresponding \bz\ and \nz\ values are $-139 \pm 120$\,G and
  $-220 \pm 140$\,G, respectively.  }
\end{figure}
\begin{figure}
\rotatebox{0}{\scalebox{0.60}{
\includegraphics*[3.5cm,5.5cm][21cm,22.5cm]{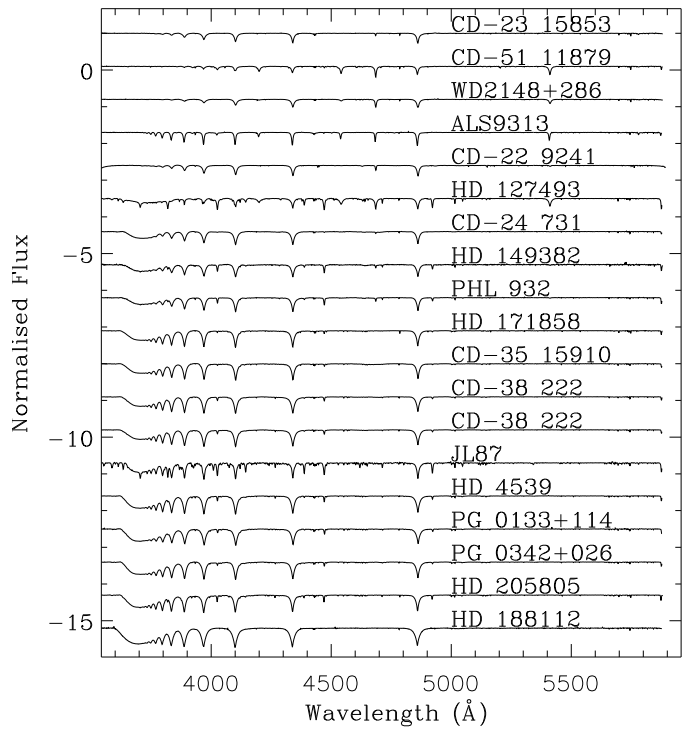}}}\\
\caption{\label{Fig_All_I} 
The normalised fluxes of 18 stars of Table~\ref{Tab_Observations}
for which new field determinations are presented in this paper.
}
\end{figure}

Polarised spectra of 17 hot subdwarfs were obtained in service mode
between March and September 2005 (one star was observed twice) with
grism 600\,B and a 0.5\arcsec\ slit width, for a spectral resolution
between 1400 and 1500. Data were reduced and analysed as described in
\citet{bagnuloetal12}, with the difference that we have implimented a
sigma-clipping process in the determination of the magnetic field from
the correlation diagram of circular polarisation against local flux
derivative. The effective Land\'e factor was set to 1 for H Balmer
lines and 1.15 everywhere else. The observing log and results are
summarised in Table~\ref{Tab_Observations}.  Successive columns list
two star names, spectral type, stellar parameters mostly found from
the hot subdwarf database \citep[][see
http://catserver.ing.iac.es/sddb/searchcat.html]{ostensenetal10}, the
exposure times and the peak signal-to-noise ratio per \AA\ in the
spectrum, the MJD of mid-observation, and the field strength \bz\ and
associated standard error in G as inferred from Stokes\,$V$. Null
profiles \citep[which are representative of the error of the observed
Stokes profiles, see, e.g.,][]{bagnuloetal09} were also calculated,
and the mean longitudinal field obtained from the null profile was
always found consistent with zero within the error bars. A thorough
discussion on this kind of diagnostic check is reported in
\citet{bagnuloetal12}.  An example of reduced data is shown in
Figure~\ref{Fig_HD205805}.

Table~\ref{Tab_Observations} includes the six field measurements
discussed by \citet{otooleetal05} and revised for the present work. Note
that the actual field strengths deduced from the FORS1 data for the
stars observed by \citet{otooleetal05} are slightly (but not
significantly) different from the revised estimate published by
\citet{bagnuloetal12} because the implementation of the sigma-clipping
algorithm has marginally changed the output of our reductions, and
because of slightly different choices of the effective Land\'{e} factors
adopted for He and metal lines.

 Two observations of the sdO star CD-22~9142, originally taken for
a white dwarf observing project \citep{jordanetal07}, were also
included in the table. One additional archive measurements for
\object{WD~0958$-$073} was reduced, but not included in
Table~\ref{Tab_Observations} because the measurement, although a null,
has a standard error of about 5~kG.

Figure~\ref{Fig_All_I} shows the normalised Stokes~$I$ spectra for all
newly measured stars of Table~\ref{Tab_Observations}.

None of the 26 field measurements of 24 different  sdB and sdO
stars shows any significant detection. (Although two measurements, of
HD~4539 and PG~0133+144, have \bz\ values that differ from zero by
somewhat more than $2\,\sigma$, this does not strongly suggest the
presence of fields, as \citet{bagnuloetal12} have shown that even with
the new reductions, standard errors of FORS1 field measurements tend
to be somewhat underestimated.)  This null result is confirmed with
other reasonable choices for reduction flags, which \citep[as
discussed by][]{bagnuloetal12} change the final field values from the
values tabulated here by typically $1\,\sigma$. Four measurements
reported in Table~\ref{Tab_Observations} are re-observations of
previously observed hot subdwarfs. In two cases the initial report of
non-detection of a field, of HD~149382 \citep{borraetal83} and of
SB~815 \citep{mathysetal12}, is confirmed by our non-detection; and
our two non-detections of a field in SB~290, together with two
non-detections in three measurements in the initial study by
\citet{mathysetal12}, suggest that the single marginal detection
reported of this star is probably not real.

 One of the original aims of these observations was to investigate the
magnetic properties of hot subdwarfs from different populations
(He-rich or in binaries for example); however, the non-detections
reported here suggest that if there are differences, they are too
small for us to currently detect.

\section{Discussion and conclusions}
Considering our own data in Table~\ref{Tab_Observations}, together with the
measurements reported in the literature, we conclude that there are
{\em no secure detections} of magnetic fields in hot subdwarf stars yet,
although at least three or four hot subdwarfs have reported fields that
may be shown on further study to be real. Looking at the more than
twenty stars discussed here, for which field measurements have failed
to reveal any field with standard errors of the order of 2--400~G, we
conclude that globally dipolar fields of kG strength occur in {\em at
  most} a few percent of hot subdwarfs. Our data are consistent with
the possibility that fields so large may be completely absent from the
hot subdwarfs.  This result suggests that the most valuable kind of
further observational survey would be one which reaches the smallest
meaurement uncertainty possible for a significant number of hot
subdwarfs. 

The apparent magnitudes of hot subdwarfs, and the richness of their
spectra, varies greatly from one star to another, but many are
suitable for field measurements with uncertainties of the order of a
hundred G or even a few tens of G. This precision can be reached
with a variety of spectropolarimeters, including some of those
discussed in Sect.~2. We believe that it is quite important to carry
out substantial further surveys for hot subdwarf fields, similar to the
ones carried out by \citet{otooleetal05} and \citet{mathysetal12}, and
particularly to re-observe hot subdwarfs for which field detections have
been reported and not later shown to be spurious.

Such surveys of the brighter hot subdwarfs, especially those with
numerous fairly sharp spectral lines, are best carried out on a
high-resolution spectropolarimeter on intermediate-size telescopes,
such as ESPaDOnS at CFHT, Narval at the Observatoire du Pic du Midi,
or HARPSPol at ESO, La Silla.  However, for fainter objects or for
ones whose spectra are dominated by broad lines (e.g. Balmer lines),
low-resolution spectropolarimeters, especially those on large-aperture
telescopes, such as ESO's FORS2, the UAGS spectropolarimeter at the
Russian 6-m telescope, or the Steward Observatory instrument have an
important and valuable role to play.

\begin{acknowledgements}

  JDL acknowledges financial support from the
  Natural Sciences and Engineering Research Council of Canada.

\end{acknowledgements}

\bibliographystyle{aa}
\bibliography{sd.bib}

\begin{thebibliography}{13}
\expandafter\ifx\csname natexlab\endcsname\relax\def\natexlab#1{#1}\fi

\bibitem[{{Bagnulo} {et~al.}(2009){Bagnulo}, {Landolfi}, {Landstreet}, {Landi
  Degl'Innocenti}, {Fossati}, \& {Sterzik}}]{bagnuloetal09}
{Bagnulo}, S., {Landolfi}, M., {Landstreet}, J.~D., {et~al.} 2009, \pasp, 121,
  993

\bibitem[{{Bagnulo} {et~al.}(2012){Bagnulo}, {Landstreet}, {Fossati}, \&
  {Kochukhov}}]{bagnuloetal12}
{Bagnulo}, S., {Landstreet}, J.~D., {Fossati}, L., \& {Kochukhov}, O. 2012,
  \aap, 538, A129

\bibitem[{{Borra} {et~al.}(1983){Borra}, {Landstreet}, \&
  {Thompson}}]{borraetal83}
{Borra}, E.~F., {Landstreet}, J.~D., \& {Thompson}, I. 1983, \apjs, 53, 151

\bibitem[{{Elkin}(1996)}]{elkin96}
{Elkin}, V.~G. 1996, \aap, 312, L5

\bibitem[{{Jordan} {et~al.}(2007){Jordan}, {Aznar Cuadrado}, {Napiwotzki},
  {Schmid}, \& {Solanki}}]{jordanetal07}
{Jordan}, S., {Aznar Cuadrado}, R., {Napiwotzki}, R., {Schmid}, H.~M., \&
  {Solanki}, S.~K. 2007, \aap, 462, 1097

\bibitem[{{Kawka} {et~al.}(2007){Kawka}, {Vennes}, {Schmidt}, {Wickramasinghe},
  \& {Koch}}]{kawkaetal07}
{Kawka}, A., {Vennes}, S., {Schmidt}, G.~D., {Wickramasinghe}, D.~T., \&
  {Koch}, R. 2007, \apj, 654, 499

\bibitem[{{Landstreet}(2004)}]{land05}
{Landstreet}, J.~D. 2004, in IAU Symposium, Vol. 224, The A-Star Puzzle, ed.
  {J.~Zverko, J.~Ziznovsky, S.~J.~Adelman, \& W.~W.~Weiss}, 423--432

\bibitem[{{Mathys} {et~al.}(2012){Mathys}, {Hubrig}, {Mason}, {Michaud},
  {Sch{\"o}ller}, \& {Wesemael}}]{mathysetal12}
{Mathys}, G., {Hubrig}, S., {Mason}, E., {et~al.} 2012, Astronomische
  Nachrichten, 333, 30

\bibitem[{{{\O}stensen} {et~al.}(2010){{\O}stensen}, {Oreiro}, {Solheim},
  {Heber}, {Silvotti}, {Gonz{\'a}lez-P{\'e}rez}, {Ulla}, {P{\'e}rez
  Hern{\'a}ndez}, {Rodr{\'{\i}}guez-L{\'o}pez}, \& {Telting}}]{ostensenetal10}
{{\O}stensen}, R.~H., {Oreiro}, R., {Solheim}, J.-E., {et~al.} 2010, \aap, 513,
  A6

\bibitem[{{O'Toole} {et~al.}(2005){O'Toole}, {Jordan}, {Friedrich}, \&
  {Heber}}]{otooleetal05}
{O'Toole}, S.~J., {Jordan}, S., {Friedrich}, S., \& {Heber}, U. 2005, \aap,
  437, 227

\bibitem[{{Petit} {et~al.}(2011){Petit}, {Van Grootel}, {Bagnulo}, {Charpinet},
  {Wade}, \& {Green}}]{petitetal11}
{Petit}, P., {Van Grootel}, V., {Bagnulo}, S., {et~al.} 2011, ArXiv:1110.5227

\bibitem[{{Savanov} {et~al.}(2011){Savanov}, {Romanyuk}, {Semenko}, \&
  {Dmitrienko}}]{savanovetal11}
{Savanov}, I.~S., {Romanyuk}, I.~I., {Semenko}, E.~A., \& {Dmitrienko}, E.~S.
  2011, Astronomy Reports, 55, 1115

\bibitem[{{Valyavin} {et~al.}(2006){Valyavin}, {Bagnulo}, {Fabrika},
  {Reisenegger}, {Wade}, {Han}, \& {Monin}}]{valyavinetal06}
{Valyavin}, G., {Bagnulo}, S., {Fabrika}, S., {et~al.} 2006, \apj, 648, 559

\end{thebibliography}

\end{document}